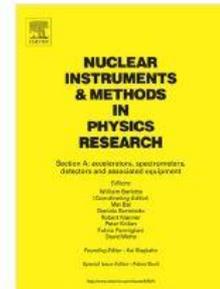

# Novel Silicon Photomultipliers suitable for Dual-Mirror Small-Sized Telescopes of the Cherenkov Telescope Array

G. Romeo[a*], G. Bonanno[a], G. Sironi[b], M.C. Timpanaro[a]

[a]*INAF, Osservatorio Astrofisico di Catania, Via S. Sofia 78, I-95123 Catania (Italy)*
[b]*INAF Osservatorio Astronomico di Brera, via Emilio Bianchi 46, 23807, Merate (LC), (Italy)*

**Abstract**

Many of the characteristics of Silicon Photomultipliers (SiPMs), such as high Photon Detection Efficiency (PDE), are well matched to the requirements of the cameras of the Small-Sized Telescopes (SSTs) proposed for the Cherenkov Telescope Array. In fact, compared to a single mirror, the double mirror Schwarzschild-Couder configuration provides a much better Point Spread Function over a large field of view. It allows better correction of aberrations at large off-axis angles and facilitates the construction of compact telescopes. Moreover, the small plate scale of the dual-mirror SSTs allows the use of SiPM detectors despite their small pixel sizes. These sensors have two further advantages compared to the Photo Multipliers Tubes: the low cost and the possibility to observe in very high Night Sky Background (NSB) light level without any damage. However, one area in which SiPM performance has required improvement is Optical Cross-Talk (OCT), where multiple avalanches are induced by a single impinging photon. OCT, coupled with the typical NSB rate of 25 MCnts/s per pixel during Cherenkov observations, can place severe constraints on the triggering capability of the cameras. This paper describes the performance of novel Low Voltage Reverse (LVR) $2^{nd}$ and $3^{rd}$ generation Multi-Pixel Photon Counters manufactured by Hamamatsu Photonics. These are designed to have both enhanced PDE and reduced OCT. Two $7 \times 7$ mm$^2$ S14520 LVR2 MPPCs with 75 μm micro-cells are tested and compared with detectors of the same pixel size with 50 μm micro-cells. A comparative analysis of a $3 \times 3$ mm$^2$ S14520 LVR2 device and an S14520 LVR3 device is also carried out, demonstrating that the LVR3 gives better photon detection in the $240 - 380$ nm wave-length range. Finally, the effect of an infrared filter on the OCT is analysed.

## 1. Introduction

The Cherenkov Telescope Array (CTA) [1] is the next generation ground-based observatory for gamma-ray astronomy at very-high energies. It will be based on more than one hundred telescopes, located in two sites (in the northern and southern hemispheres). CTA will build on the strengths of current Imaging Atmospheric Cherenkov Telescopes (IACT), employing three telescope designs combined in the Large-Sized Telescope, Medium-Sized Telescope, and Small-Sized Telescope (SST) array to enhance the sensitivity by up to an order of magnitude compared with current facilities [1]-[2] in the 100 GeV to 10 TeV range and extend the accessible energy range from 20 GeV to 300 TeV. The SST array sensitivity and coverage are optimized from a few TeV to 300 TeV. About 70 SST telescopes are planned to be placed in the CTA southern site to cover several square kilometres in order to achieve the necessary sensitivity. Because of the large number of telescopes required, cost reduction is critical for the SST. However, the camera cost cannot be reduced by simply decreasing its size, due to the minimum number of pixels required (>1000) and the relatively high unit cost of traditional photon sensing photomultiplier tubes (PMTs). In order to mitigate this problem, the SST utilizes silicon photomultipliers (SiPMs) as the baseline photon sensor technology [3]. Currently, SiPM detectors with large size (i.e. $20 \times 20$ mm$^2$) surely suffer from high dark count rates and very high recovery time. One solution is to make use of dual-mirror optics which have a relatively small plate scale and whose focal plane can be covered with SiPMs with sizes ranging from $6 \times 6$ mm$^2$ to $7 \times 7$ mm$^2$ [4].

SiPM detectors, thanks to their outstanding characteristics in terms of photon detection efficiency, photon number resolution, low operating voltage, fast dynamic response and insensitivity to magnetic fields, are suitable in the fields of high-energy astrophysics and IACT applications [5]-[13]. Considerable effort is presently being invested by the producers of SiPMs to further improve the performance achieved by this class of devices [14]-[16]; moreover, characterization studies and methodologies for evaluating the detector performance have been carried out [17]-[25]. Currently, we can safely assert that the SiPMs Photon Detection Efficiency (PDE) is greater than that of PMTs in the 300 - 700 nm spectral range, but another SiPM parameter has to be carefully considered: the optical cross-talk, a mechanism whereby a single optical photon can produce multiple avalanches in the SiPM. In CTA, the night sky background level of typically 25 MCnts/s/pixel places severe constraints on the trigger capability due to accidental coincidence of neighbouring pixels. In order to suppress such events, it is necessary to reduce optical cross-talk at very low level, while keeping good Photon Detection Efficiency. An OCT probability lower than 10% and a PDE higher than 25% meets the SST requirements.

* *Corresponding author*: Giuseppe Romeo (giuseppe.romeo@inaf.it)

This paper presents the characterization of a newly available large-area Low Voltage Reverse (LVR) $2^{nd}$ and $3^{rd}$ version Multi-Pixel Photon Counter (MPPC) detectors manufactured by Hamamatsu Photonics (HPK). These devices, can reach saturating PDE at lower over-voltage compared with Low Cross Talk (LCT), which means LVR devices can achieve higher PDE at lower OCT since OCT scales with over-voltage. Such detectors are suitable for the camera focal plane of the SST double mirror (Astrofisica con Specchi a Tecnologia Replicante Italiana) ASTRI telescopes [8] proposed for the CTA southern site. In order to show the perfect ASTRI camera fitting and the good geometrical fill factor achieved, a brief description of the relevant mechanical parameters is also provided.

Two $7 \times 7$ mm$^2$ S14520 LVR2 MPPCs with 75 μm micro-cells have been tested and a comparative analysis of the large pixel pitch (75 μm) detector versus the smaller pixel pitch (50 μm) device with the same active area is here described.

HPK selected the product code S14520 for the LVR devices (dedicated only to the CTA project). From now on we will cite the LVR series and leave out the product code.

We also carried out PDE measurements on a $3 \times 3$ mm$^2$ LVR3 SiPM detector and compared it with an LVR2 device. The results demonstrate that the LVR3 device gives better PDE in the 280 - 380 nm spectral range.

The effect of an optics (an infrared cut-off filter in the specific case of the ASTRI camera) placed in front of an SiPM on the OCT is described.

The measurements shown here are carried out at the Catania astrophysical Observatory Laboratory for Detectors (COLD) within INAF – Osservatorio Astrofisico di Catania.

**2. ASTRI focal plane coverage and SiPM relevant mechanical parameters**

The double-mirror optical configuration of the ASTRI telescopes permit us to design a compact and lightweight camera to be placed at the curved focal surface. The detection surface requires a spatial segmentation with an interspace of a few millimetres to be compliant with the imaging resolving angular size (0.19°). In order to match the angular resolution of the optical system, the design of the ASTRI camera has to comply with the dimensions of the single pixel and of the basic detection module (the tile). Since the convex shaped focal surface of the ASTRI camera has a curvature radius of about 1m, the curved surface of the camera has to fit with a certain number of square pixels without losing the required focusing capability of the optical system. This specification is physically accomplished by a pixel of about $7 \times 7$ mm$^2$ and a detection module of $57.6 \times 57.6$ mm$^2$.

Figure 1 shows the focal plane assembly. A total of 37 detection modules (2368 pixels) form the camera at the focal plane; this structure is capable of achieving a full Field of View (FoV) of 10.5°.

The technical feasibility of this design has been demonstrated with the ASTRI optical validation [26].

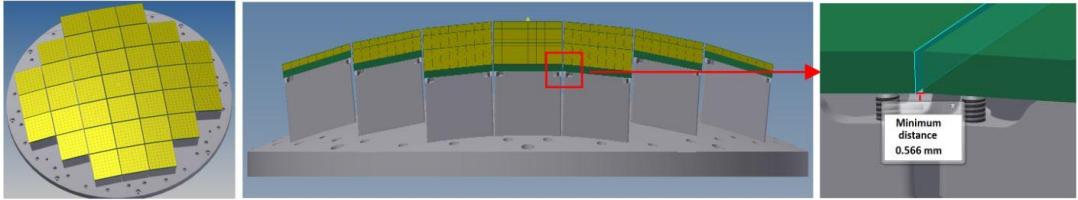

Figure 1. ASTRI camera focal plane assembly. The telescope curved focal plane imposes the arrangement of detection modules and the maximum allowable dimensions of the tiles.

Using $7 \times 7$ mm$^2$ SiPM detectors with a pitch-size of 75 μm, 93 micro-cells fill completely a pixel sensitive area, giving a total of 8649 micro-cells. This ensures a sufficiently high dynamic range as required by the CTA project. An $8 \times 8$ tile will cover a total area of $57.6 \times 57.6$ mm$^2$ meaning 3317.76 mm$^2$, $6.975 \times 6.975$ mm$^2$ being the active area of each pixel, 0.2 mm the interspace between pixels and 0.2 mm the tile edge spacing. The total active area is 3091.36 mm$^2$ and thus a tile geometrical filling factor of 93.18 % is achieved. Figure 2 shows the tile schematics, a single $7 \times 7$ mm$^2$ pixel and the edge detail.

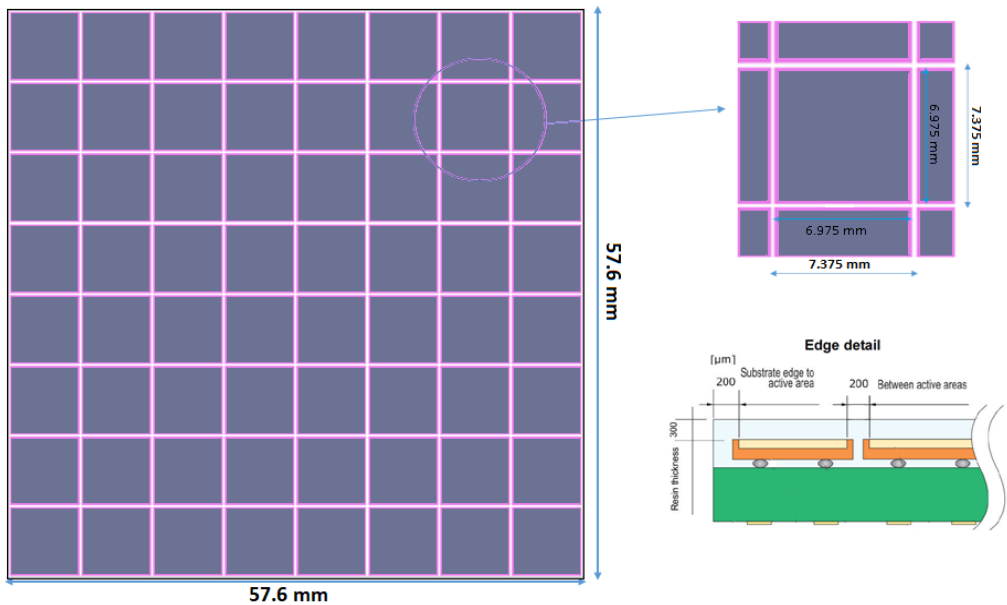

Figure 2. $8 \times 8$ pixel tile schematics. Using $7 \times 7$ mm$^2$ SiPMs $57.6 \times 57.6$ mm$^2$ can be assembled. On the right part a detail of a single $7 \times 7$ mm$^2$ pixel is shown. With this configuration a total area of 3317.76 mm$^2$, an active area of 3091.36 mm$^2$ and a geometrical filling factor of 93.18 % are achieved.

## 3. Large-Area Low Voltage Reverse 2$^{nd}$ and 3$^{rd}$ version SiPM detectors

The optical trench improvement, started with the LCT series and characterized by new types of trenches that enabled cross-talk reduction, has been continued in the new denominated Low Voltage Reverse family. On the other hand, the fill-factor improvement of the new LVR series is achieved by maximizing the active area.

The characterized large-area SiPM described in this paper belongs to the latest device series manufactured by Hamamatsu Photonics, denominated LVR family and reported as the MPPC LVR2 CS (Ceramic package Silicone coating) and CN (Ceramic package No coating) and LVR3 CS and CN series. These are prototype devices provided by the vendor to the COLD laboratory for testing and evaluation purposes. Figure 3 shows the three characterized devices while Table I reports the main features of the characterized detectors.

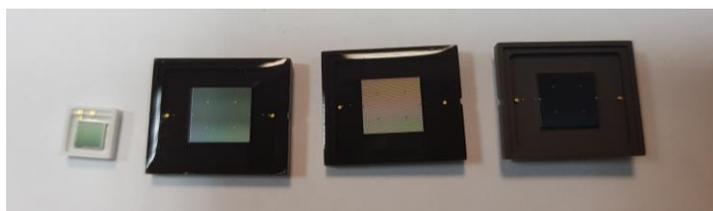

Figure 3. Characterised devices: from left to right: LVR3 3050 CN, LVR2 7050 CS, LVR2 7075 CS and LVR2 7075 CN.

TABLE I
Main physical features of the characterized MPPC detectors

| device series | S14520(LVR2 7075 CS) | S14520(LVR2 7075 CN) | S14520(LVR2 7050 CS) | S14520 (LVR3 3050 CN) |
|---|---|---|---|---|
| **cell pitch** | 75 μm | 75 μm | 50 μm | 50 μm |
| **device size** | $7 \times 7$ mm$^2$ | $7 \times 7$ mm$^2$ | $7 \times 7$ mm$^2$ | $3 \times 3$ mm$^2$ |
| **sensitive area** | $6.975 \times 6.975$ mm$^2$ | $6.975 \times 6.975$ mm$^2$ | $6.950 \times 6.950$ mm$^2$ | $3.00 \times 3.00$ mm$^2$ |
| **micro-cells** | 8649 | 8649 | 19321 | 3600 |
| **surface coating** | Silicone resin | No coating | Silicone resin | No coating |
| **cell fill-factor** | 82 % | 82 % | 74 % | 74 % |
| **breakdown voltage*** | 38.00 V | 38.60 V | 38.20 V | 38.68 V |

* at 25°C

## 4. Characterisation Setup

The electro-optical equipment used for SiPM measurements is depicted in Figure 4.

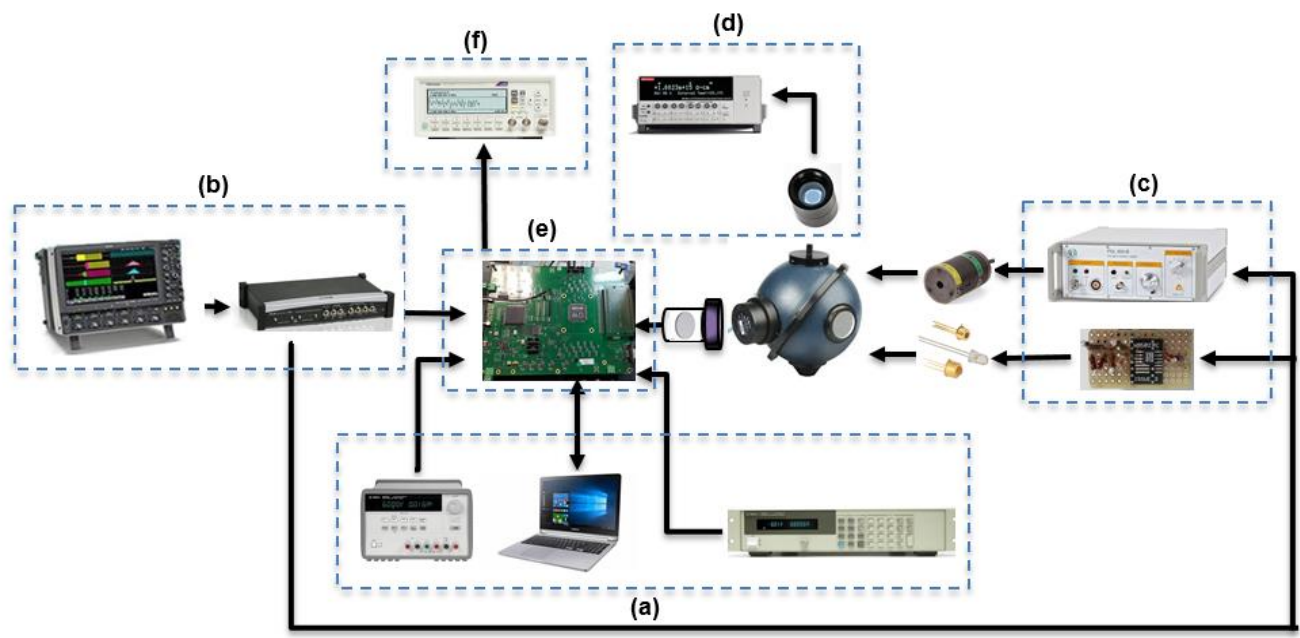

Figure 4. Characterization setup based on pulsed light sources: lasers and LEDs. Apart from the integrating sphere, that can be considered the central part of the system, 5 main blocks can be envisaged: a) power supply, b) pulse generator, c) laser and LED drivers, d) calibrated reference photodiode, e) CITIROC-1A electronic front-end, f) counter.

Along the schematics, five main blocks can be envisaged:
a) SiPM and electronic front-end power supply: made up of an Agilent 6634B to supply the high voltage to the SiPM and an Agilent E3631A to supply the voltage levels to the LED driver.
b) Pulse generator: made up of an oscilloscope LeCroy wavePro 725Zi 2.5GHz and a pulse generator LeCroy ArbStudio 1104 to generate the synchro signals for the pulsed light source.

c) Laser and LED controllers: made up of a PicoQuant PDL 200-B to drive the lasers and an appropriate electronic circuit based on an AD8009 ultra-fast amplifier, capable of conditioning signals at a slew rate of 5.5 V/μs (545 ps rise time).
d) Calibrated reference detector: made up of an IRD (International Radiation Detectors) traceable photodiode NIST (National Institute of Standards and Technology) and a Keithley 6514 electrometer to measure the photo-current.
e) CITIROC-1A electronic front-end: to drive and process the SiPM signals.
f) Counter Tektronics FCA 3000: to measure the dark count rate for the staircase

Furthermore, a set of calibrated neutral density filters ND30B, ND20B, ND13B and ND10B by Thorlabs[1] are used in front of the SiPM to attenuate the signal and avoid working with a luminous level that could saturate the detector, while having sufficient current signal detected by the photodiode.

The apparatus allows PDE measurements in the spectral range of 280 – 850 nm by using 18 pulsed light sources.

The wave-length of each source is reported in Table II.

TABLE II
Pulsed light sources and their respective wave-length available at the COLD laboratory

| Identifier code | Wave-length [nm] | Type |
|---|---|---|
| LED285W | 285 | LED |
| LED315W | 315 | LED |
| LED341W | 341 | LED |
| LED385L | 385 | LED |
| LDH-P-C-405 | 405 | Laser |
| LED430L | 430 | LED |
| EPL-450 | 450 | Laser |
| LED-450 | 450 | LED |
| LED465E | 465 | LED |
| PLS-8-2-746 | 496 | LED |
| LED505L | 505 | LED |
| LED525L | 525 | LED |
| LED570L | 570 | LED |
| LED591E | 591 | LED |
| LDH-P-635 | 635 | Laser |
| LED660L | 660 | LED |
| LED680L | 680 | LED |
| LED780E | 780 | LED |
| LED851L | 851 | LED |

Temperature control and stabilization (± 0.5°C) is obtained through a dedicated thermoelectric cooling system based on a Peltier cell; the entire cooling system is thermally calibrated and can achieve temperatures from 2°C to 30°C. The mechanical housing is able to host various types of detectors by simply using a dedicated electronic adapter board.

As stated above, the main element of the front-end electronics for SiPM characterization is the Cherenkov Imaging Telescope Integrated Read-Out Chip (CITIROC) [27], which is an advanced version of the Extended Analogue SiPM

---

[1] https://www.thorlabs.com/navigation.cfm?guide_id=2185

Integrated Read-Out Chip (EASIROC) [28], both produced by WEEROC[2]. The chip has been selected as the front-end electronics for the ASTRI camera. The analog core of the chip has 32 channels, each one incorporating:

- A digital-to-analog converter (DAC) for the SiPM high voltage adjustment.
- Two preamps that allow setting dynamic range, through variable capacitors, between 160 fC and 320 pC.
- A trigger line consisting of two fast shapers (15 ns shaping time) and two discriminators.
- Two slow shapers (from 12.5 ns to 87.5 ns shaping time) and two track and hold blocks are responsible for measurement of the Pulse Height Distribution.

The block diagram of one channel is shown in Figure 5.

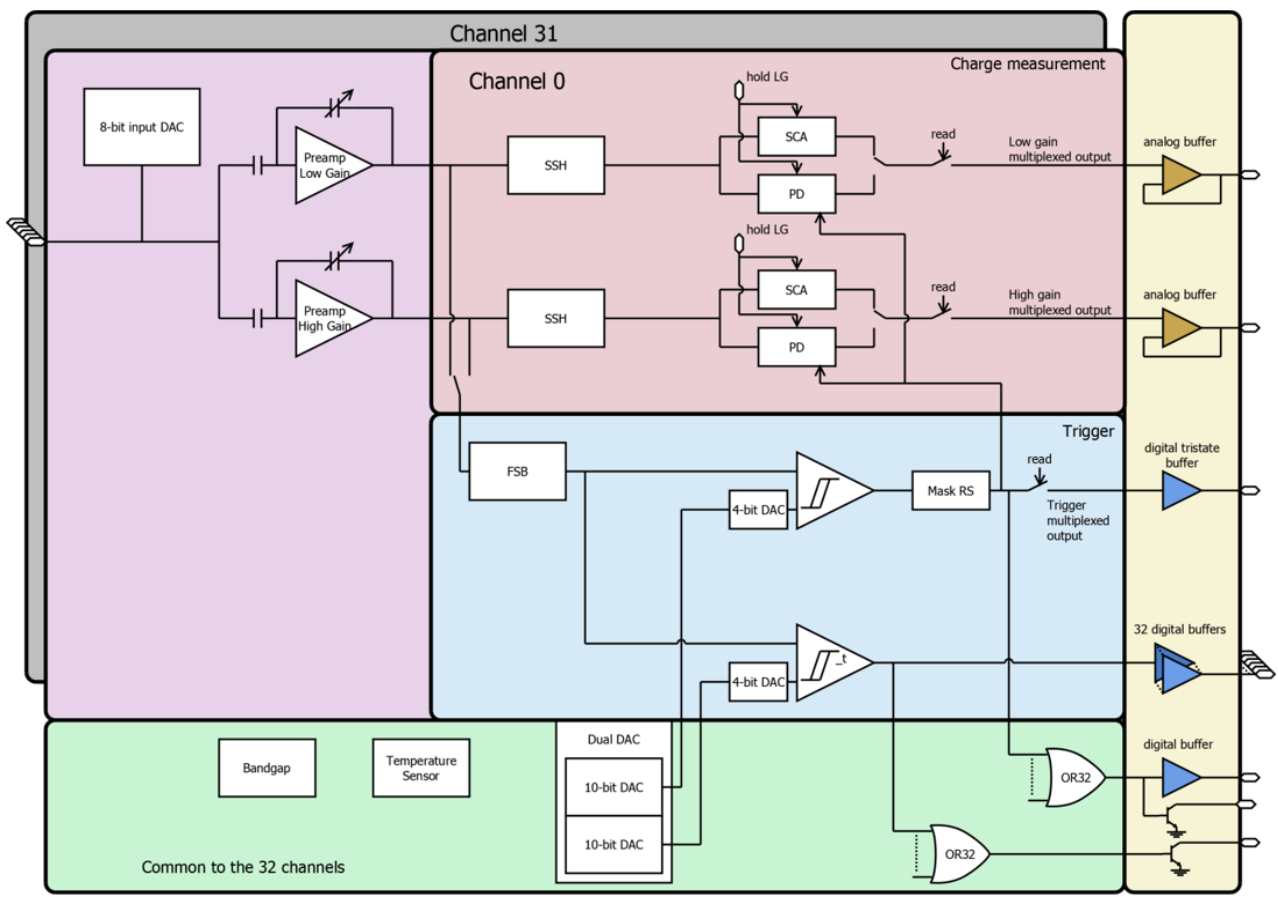

Figure 5. CITIROC-1A single channel block diagram.

The part of the ASIC chip used to obtain staircase measurements is depicted in Figure 6. The SiPM pulse is amplified by the High Gain (HG) preamplifier and then is shaped (15 ns shaping time) by the bipolar fast shaper. A discriminator with a programmable threshold (10 bit DAC) is used to drive the output signal through the OR32, and in turn, to drive the counter Tektronics FCA 3000.

---

[2] http://www.weeroc.com

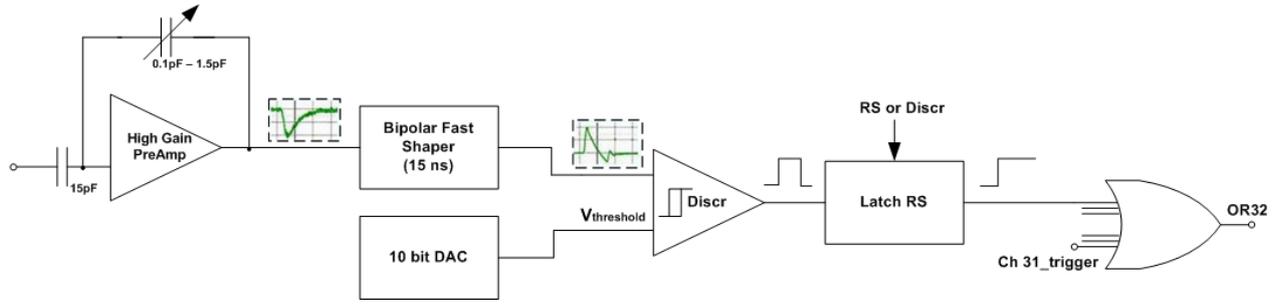

Figure 6. Block diagram of the used electronics inside the ASIC for staircase measurements.

As we will discuss later, the 15 ns shaping time of the bipolar fast shaper is not fast enough and is responsible for the pile-up of dark counts due to slow baseline recovery. Since two 1-p.e. pulses will be counted as one 2-p.e. pulse due to the pile up, the dark count rate (DCR) at 1.5 p.e. will appear higher than the real one and the DCR at 0.5 p.e. will appear lower than the real one. Of course an adequate Pole Zero Cancellation (PZC), able to operate an effective baseline recovery, will greatly mitigate the effect.

## 5. Experimental results

Based on the above mentioned equipment, experimental measurements on the LVR2 series $7 \times 7$ mm$^2$ MPPC are carried out and shown here in terms of the main detector performance parameters, i.e. Dark Count Rate, Optical Cross-Talk and Photon Detection Efficiency. In particular, the PDE has been measured in the 285 – 850 nm spectral range and at 405 nm at various operating voltages.

*A. Dark Count Rate and Optical Cross-Talk*

The DCR is essentially defined as the count rate of avalanche pulses produced by primary (uncorrelated) carriers, resulting in events that are perfectly equivalent to the signal from real photons.

SiPM optical cross-talk, as reported in literature ([17]-[20],[23]-[25]), is evaluated as the ratio of the DCR at 1.5 p.e. with respect to that at 0.5 p.e. This approach is based on the assumption that the probability of triggering two uncorrelated avalanches at the same time is negligible.

As stated in section 4, the front-end electronics suffers from a sort of a pile-up effect, that is more pronounced in large devices due to the higher DCR. In order to reduce this effect and measure a true OCT, possibly not affected by electronics, it is necessary to operate the detectors at low DCR.

As the DCR depends on temperature, we observed this sort of pile-up effect by measuring the apparent OCT at an over-voltage of 3.0 V of $3 \times 3$ mm$^2$, $6 \times 6$ mm$^2$ and $7 \times 7$ mm$^2$ LVR2 devices with operating temperature.

In Figure 7, it can be observed that, in the analysed temperature range, the OCT of the $3 \times 3$ mm$^2$ device remains at about 2.0 % in all the range, meaning that the DCR (about 300 KCnts/s at 20°C) is sufficiently low to avoid the pile-up effect, while the other two devices suffer from the effect and, at temperatures higher than 5°C (at which the DCR is 700 KCnts/s and 850 KCnts/s for the $6 \times 6$ mm$^2$ and $7 \times 7$ mm$^2$ respectively), the apparent OCT increases with temperature. Below 5°C we observe, as expected, that the apparent OCT depends only on the SiPM size and can be considered very close to the true OCT.

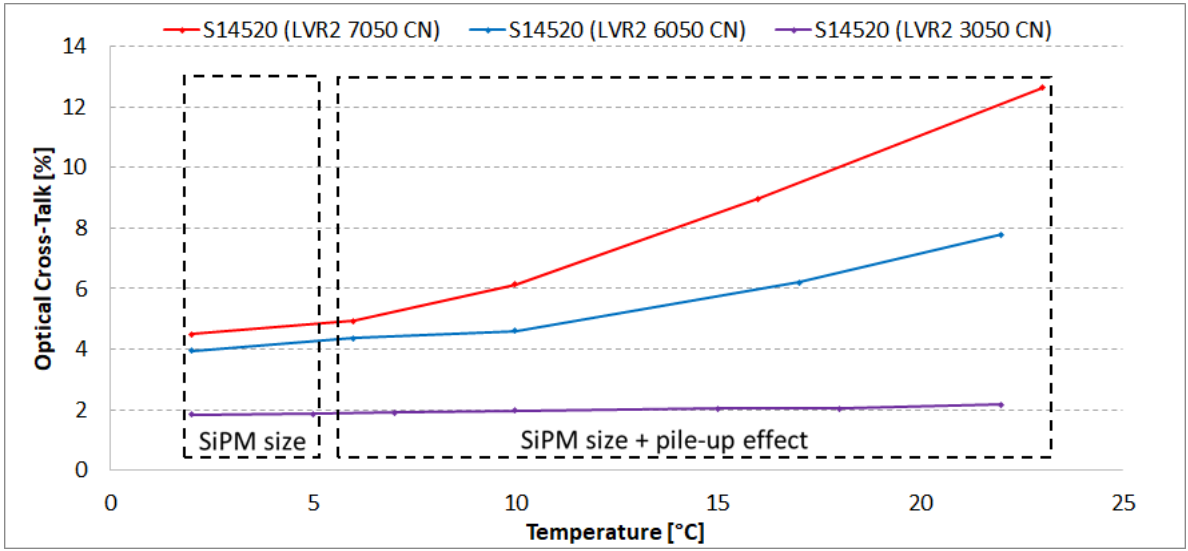

Figure 7. OCT versus temperature. Below 5°C, the OCT depends only on the SiPM size while above 5°C the OCT is affected by the pile-up effect, or better to say the electronic front-end does not operates properly.

Figure 8 shows the DCR curves (also known as staircase functions) for the characterized SiPMs LVR2 7075 CS and LVR2 7075 CN as a function of the discriminator threshold at 3°C operating temperature and for over-voltage values from 2.0 V up to 5.0 V, in steps of 1.0 V

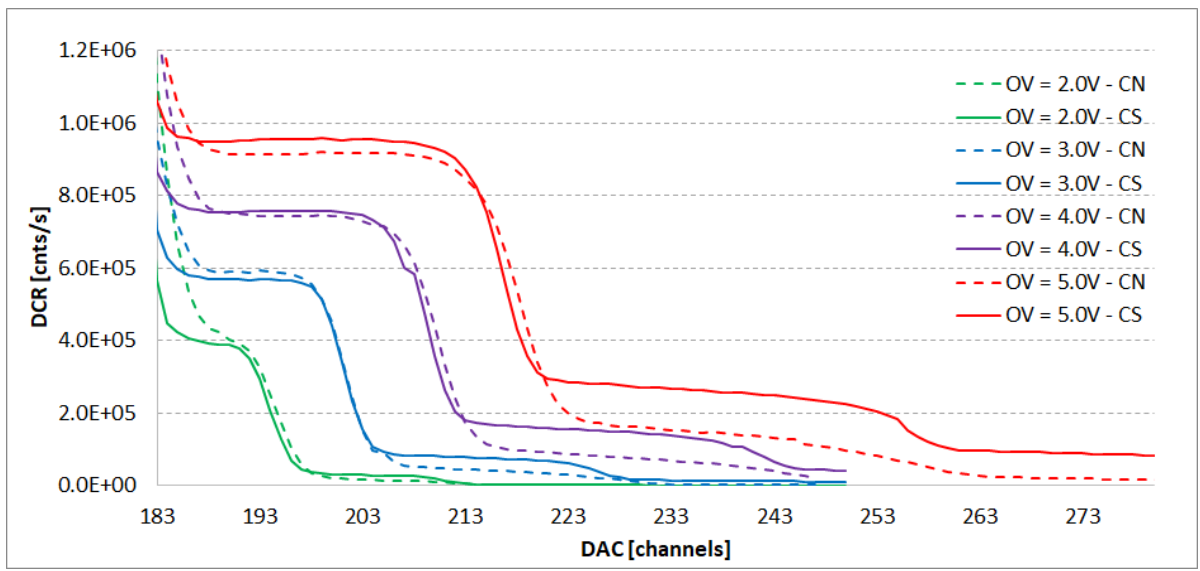

Figure 8. DCR curves for the characterized LVR2 7075 CS and LVR2 7075 CN detectors as a function of the discriminator threshold at 3°C operating temperature. The difference in the second p.e. is clearly evident and implies a reduced OCT in the CN series device.

From the obtained staircases depicted in Figure 8 we derived the OCT plots for both CS and CN devices and shown in Figure 9. In this figure we also report the OCT versus over-voltage at 15°C because this is the ASTRI camera working temperature. From measurements, we can derive the degradation of apparent OCT when we operate the camera at 15°C.

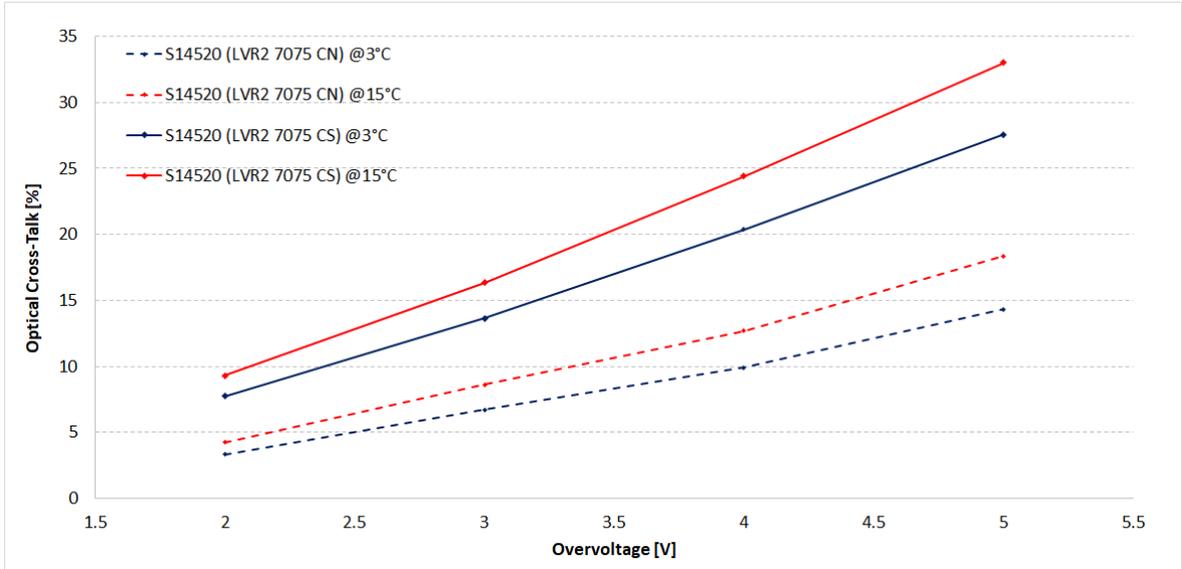

Figure 9. OCT versus over-voltage at 3°C and at 15°C. The CN series shows a lower OCT compared to the CS series because of the protective coating in the CS series. The electronics is responsible for the different behaviour at the two temperatures.

As expected and easily understood from the staircases of Figure 8, the CN series shows a lower OCT than the CS series. This is essentially due to the protective resin coating on top the SiPM sensitive surface that causes a back reflection of the photons produced during an avalanche. A higher apparent OCT at 15°C is observed because of the pile-up effect. This last can be evaluated considering that at lower temperatures (meaning lower DCR) the OCT is less affected by the pile-up. Being the OCT 7.0 % at 3°C (DCR about 0.55 MCnts/s) and 9.0 % at 15°C (DCR about 1.3 MCnts/s), the pile-up effect can be derived considering the relation between the true and the apparent OCT [30]:

$$OCT_{tr} = 1 + \left[\frac{OCT_{app} - 1}{e^{(-DCR*\Delta t)}}\right]$$

The pile-up effect is given by:

$$PU = 1 - e^{(-DCR*\Delta t)}$$

Where:

$OCT_{tr}$ is the true OCT, $OCT_{app}$ is the apparent OCT which includes pile-up effects, DCR is the dark count rate and $\Delta t$ is the two pulse resolving time.

At 3°C, $OCT_{app}$ ~ 0.07 and the DCR is 0.55 MCnts/s, while at 15°C, $OCT_{app}$ ~ 0.09 and the DCR is 1.3 MCnts/s. Imposing that the $OCT_{tr}$ and $\Delta t$ at both temperatures are the same:

$$1 + \frac{(0.07 - 1)}{e^{(-0.55*\Delta t)}} = 1 + \frac{(0.09 - 1)}{e^{(-1.3*\Delta t)}}$$

we can get $\Delta t$ = 0.029 µs and $OCT_{tr}$ = 0.055, and the pile-up effect would be:

$$PU = 1 - e^{(-DCR*\Delta t)} = 1 - e^{(-1.3*0.029)} = 0.037$$

Thus, we can conclude that the true OCT is ~ 5.5 % and the pile-up effect is ~ 3.7% at 15°C and ~ 1.5 % at 3°C.

The OCT of the SiPM LVR2 7075 CS is about double compared to that of the CN series, while an OCT difference of about 2.0 % is observed at an operating temperature of 15°C. At 3.0 V of over-voltage the OCT is 15 % while, at the same over-voltage, Low Cross Talk 5[th] generation (LCT5) CS devices showed an OCT of 21% [29]. Thus a decrease of 6% is demonstrated. The LVR2 technology has a lower break down voltage than that of the LCT5 which means a thinner avalanche region and a higher gain at the same overvoltage. Despite the higher gain that can cause higher OCT, the thinner avalanche region allows trenches optimization and OCT decrease.

*B. Photon Detection Efficiency*

SiPM absolute PDE measurements are based on the statistical analysis of the pulse heights distribution of the SiPM devices in both light and dark conditions. Thanks to these measurements, the number of pulses per unit time in monochromatic light conditions are compared to the light level recorded at the same time, by a reference NIST photodetector. The used technique was discussed in detail in [21].

The adopted photon counting technique for determining the detector PDE is insensitive to the optical cross-talk, as we measure the probability of detecting 0 photons and derive the average assuming the Poisson statistics. The average number of photons revealed by the SiPM is given by:

$$\bar{N}_{SiPM} = -ln\left[\frac{N_{ped}}{N_{ped}^{DK}}\right]$$

Where:

$N_{ped}$ is the number of events at 0 p.e. (i.e. the pedestal) during pulsed light measurements.

$N_{ped}^{DK}$ is the number of events at 0 p.e. (i.e. the pedestal) without light source (dark).

The average number of photons revealed by calibrated photodiode $\bar{N}_{ph\ cal\text{-}Diode}$ is given by:

$$N_{ph\ Cal-Diode} = \frac{N_e}{QE(\lambda)}$$

Where $N_e$ is the number of photoelectrons detected by the calibrated photodiode and $QE(\lambda)$ its Quantum Efficiency.

The PDE is then obtained by the ratio:

$$PDE = \frac{\bar{N}_{SiPM}}{\bar{N}_{ph\ Cal-Diode}} \times \frac{A_{Diode}}{A_{SiPM}}$$

Where $A_{Diode}$ and $A_{SiPM}$ are the areas of both detectors because we don't use diaphragms.
However, it has to be observed that the obtained PDE is not immune to extra-charge noise (delayed cross-talk and afterpulsing), and therefore it may be slightly overestimated.
We measured the PDE as a function of the over-voltage for the wave-length of 405 nm. To check the improvement of the LVR2 7075 CS and CN in relation to the SiPM of same series with 50 μm micro-cell, we also measured the PDE versus the over-voltage of the LVR2 7050 CS and CN. Figure 10 shows the compared PDEs.

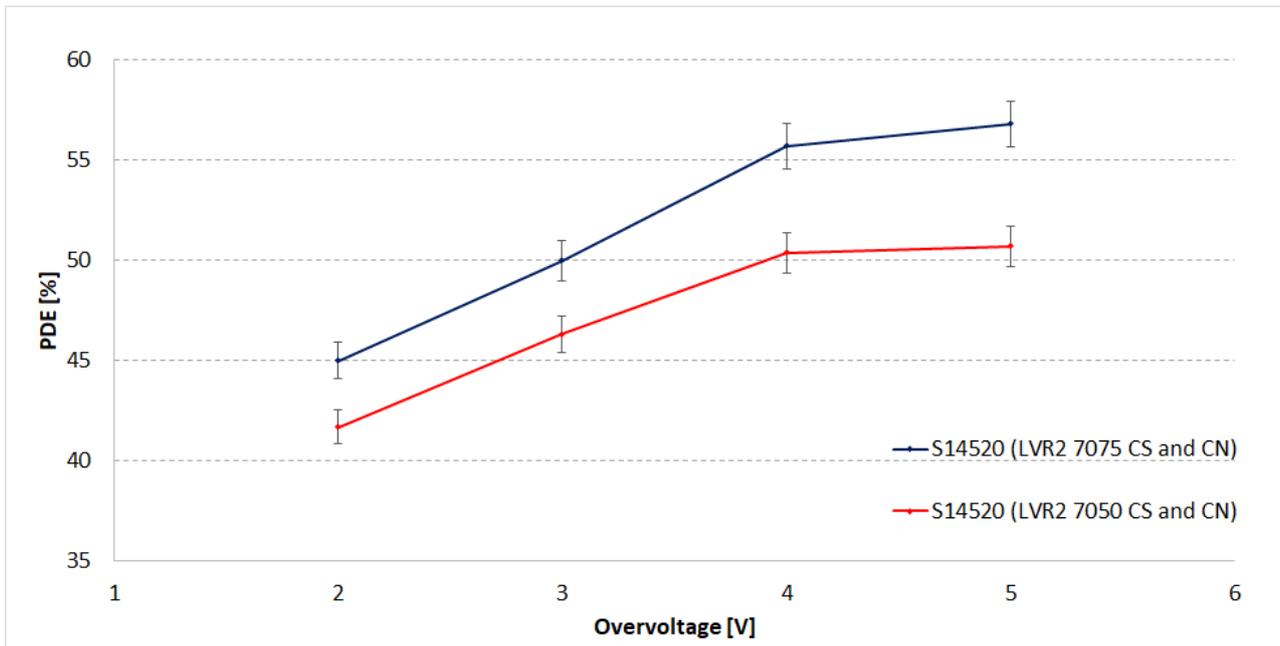

Figure 10. PDE versus over-voltage at 405 nm. LVR2 7075 CS and CN are compared to LVR2 7050 CS and CN.

As it can be noted, the LVR2 7075 CS and CN reach the PDE saturation of about 57 % at an over-voltage of 5.0 V, while the LVR2 7050 CS and CN saturate at 51 % at the same over-voltage; the difference of about 5 % is maintained at all over-voltages in the 2.0 - 5.0 V range.

For the LVR2 7075 CS, at an over-voltage of 3.0 V, a PDE of 50 % is measured, while at the same wave-length and over-voltage a PDE of 45 % is achieved for the LCT5 devices [29]. As stated in subsection A, at the same over-voltage the LVR2 device has a higher gain than the LCT5 detector, and this is responsible for a trigger probability improvement.

PDE measurements in the 310 – 850 nm wave-length range of the LVR2 7075 CN and LVR2 7075 CS have been obtained by using the pulsed sources (lasers and LEDs) described in section 4 and reported in Figure 11. The measurements have been carried out by operating the detectors at over-voltages of 3.0 V and 4.0 V; moreover, the PDE of the LVR2 7050 CS and CN has been measured and reported on the plot to better evaluate the achieved performance of the SiPM with larger micro-cells.

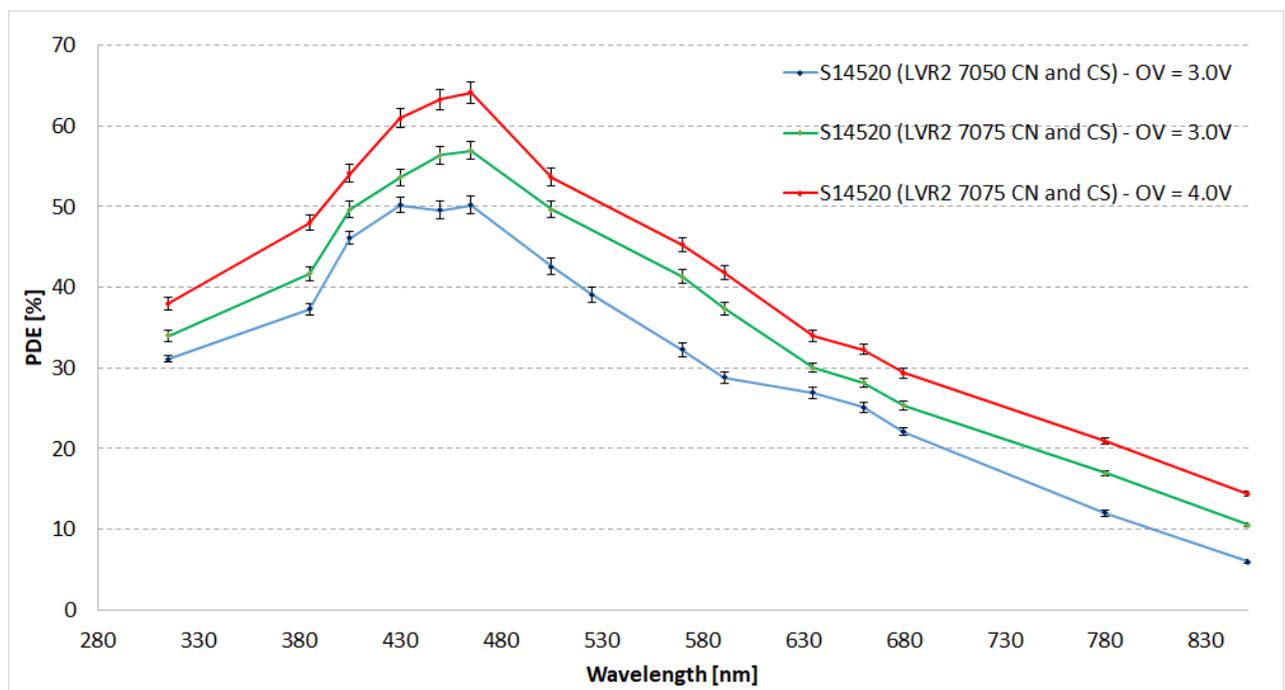

Figure 11. *PDE* measurements of the characterized LVR2 7075 CS and LVR2 7075 CN detectors at an over-voltage of 3.0 V and 4.0 V are carried out in the 310 – 850 nm spectral range. The PDE of LVR2 7050 CS and CN at an over-voltage of 3.0V is also added for comparison. A PDE of 57% at an over-voltage of 3.0 V is achieved at wave-lengths in the range 450 – 470 nm. A PDE of 64% at an over-voltage of 4.0 V, where a reasonable OCT is measured.

In the spectral range 450 - 470 nm a PDE of 57% at an over-voltage of 3.0 V is achieved. A PDE difference of about 4% between the two kinds of SiPM 50 μm and 75 μm micro-cell pitch is observed. At an over-voltage of 4.0 V, where the avalanche trigger probability is about 99%, a PDE of 64% compatible with a fill factor of 82% and a quantum efficiency of about 80%, has been measured. The impressive PDE and the relatively low OCT lay the base for a suitable $7 \times 7$ mm$^2$ SiPM for the CTA SST dual mirror telescope.

In order to better quantify the detectors performance, PDE measurements at 405 nm as a function of OCT (at an operating temperature of 3°C) for both CN and CS series devices are depicted in Figure 12.

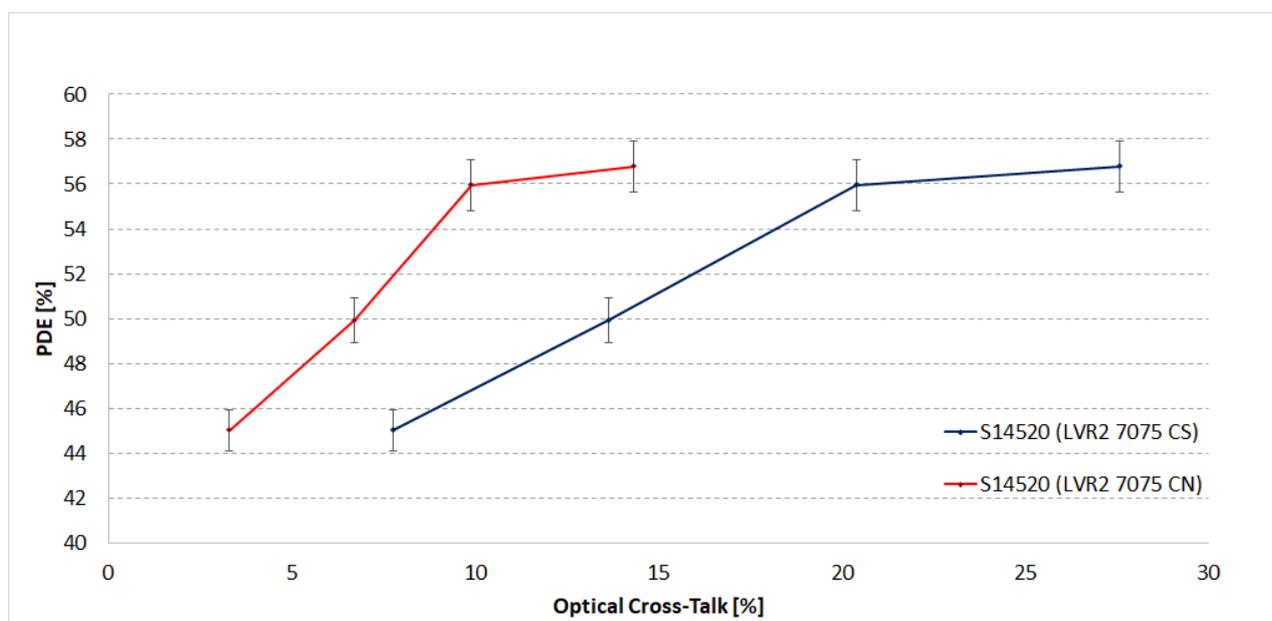

Figure 12. PDE(405nm) versus OCT for the LVR2 7075 CS and LVR2 7075 CN detectors. As it can be noted, a PDE of about-50% is achievable with only 6.5% of OCT in the CN series MPPC and with 13.5% of OCT in the CS series detector.

Such plots allow an evaluation of the optimal trade-off between PDE and optical cross-talk for the specific application. Data points in Figure 12 refer to the same over-voltage values as in Figure 10.

As it can be observed, at over-voltages of 3.0 V, a PDE of about 50% is achieved with an OCT of only 6.5% in the CN series SiPM and 13.5% in the CS series detector. As discussed in the previous subsection, the LVR2 7075 CS SiPM device gives greater optical cross-talk due to the coating, at an over-voltage of 2.0 V the OCT can be reduced to about 7.0% but at the expenses of the PDE that drops to about 45%. In applications where an OCT of ∼15% is acceptable, a PDE of ∼50% at 405nm can be achieved by the CS series device while the CN series SiPM is preferable where low cross-talk values are required. Of course CN devices increase breakage risks and need careful handling.

C. *Relative PDE of the LVR3 compared to the LVR2*

HPK made an effort to also improve the SiPM response in the 300 – 350 nm spectral range. They produced another series named LVR3 that takes into account the improvement. We received only a $3 \times 3$ mm$^2$ with 50 μm micro-cell LVR3 3050 CN. We measured the PDE of this device and compared it with that of an LVR2 device (same dimensions and micro-cell). Figure 13 shows the PDE versus wave-length in the 285 – 850 nm range obtained by operating both devices at an over-voltage of 3.0 V. A peak in PDE of about 50% is measured in the 430 – 450 nm range for both devices while, at 285 nm, the S14520 LVR3 and the S14520 LVR2 achieve a PDE of 24% and 20%, respectively. A smaller improvement is achieved at 315 nm and 385 nm.

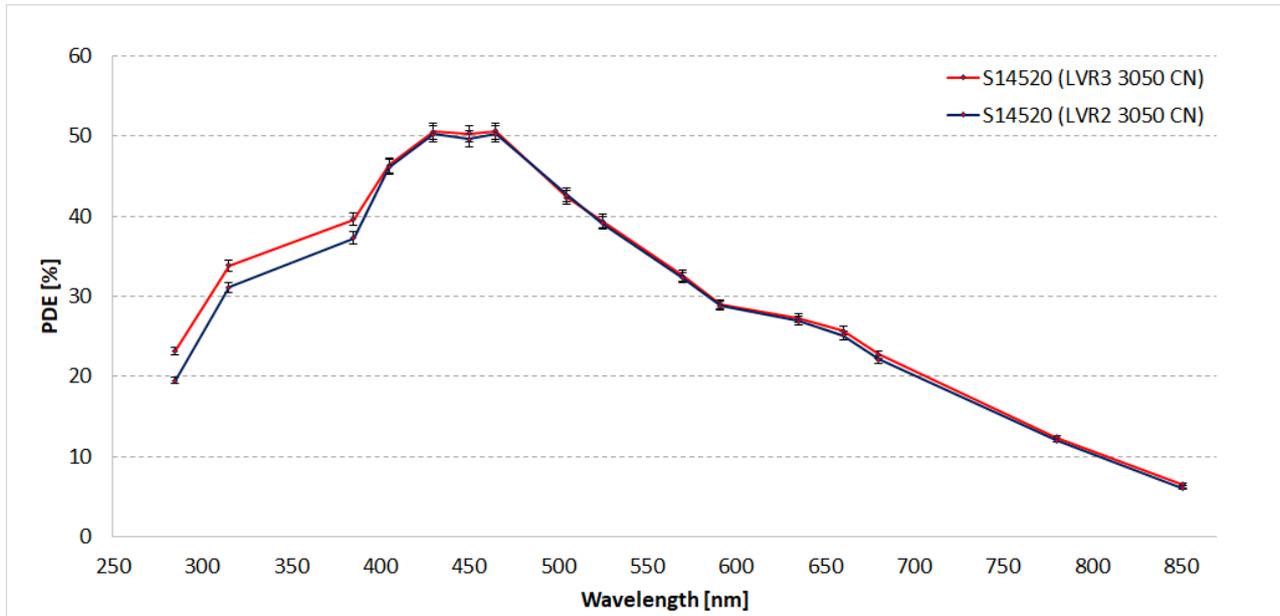

Figure 13. Superimposed LVR2 and LVR3 PDE curves. The difference in the near UV region is clearly evident

To emphasize the difference between the two technologies, we plotted the ratio of the PDE values obtained for the two devices in Figure 14. This ratio is the relative PDE.

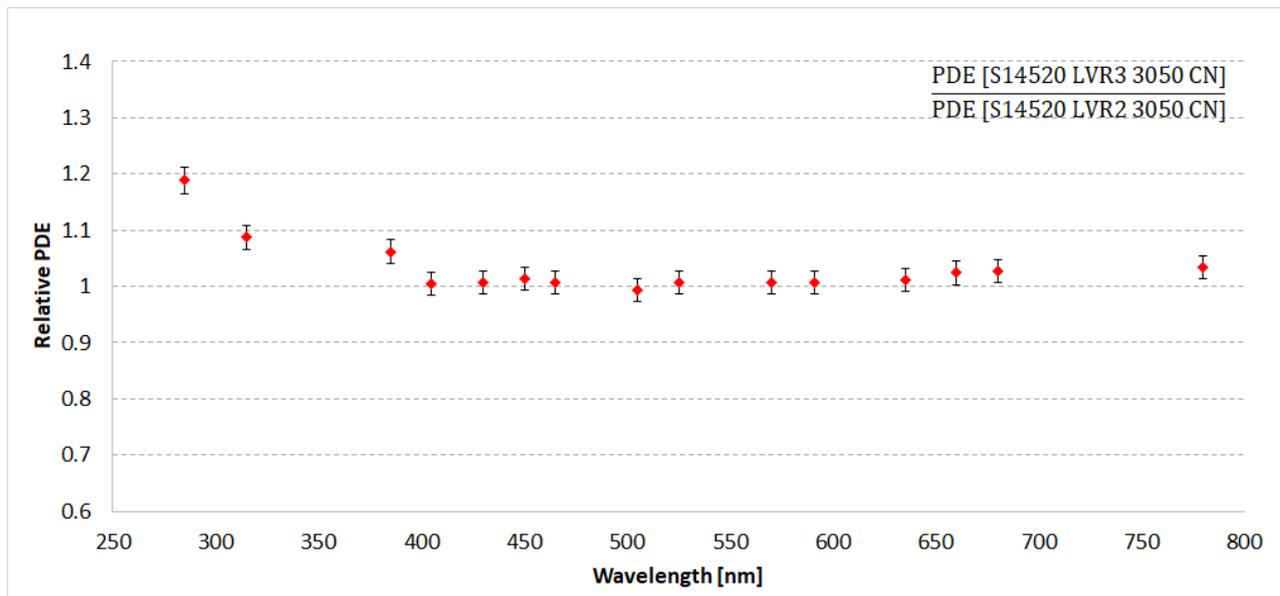

Figure 14. Relative PDE of the LVR3 3050 CN in relation to the LVR2 3050 CN.

## 6. Effect of the ASTRI IR Filter on the Optical Cross-Talk

As stated above, the ASTRI telescope focal plane is populated with SiPM detectors that have a very good sensitivity in the blue spectral range (where the observed Cherenkov light emission peaks). Unfortunately, SiPM detectors, being essentially silicon sensors, are also sensitive in the infrared energy band where the emission of the NSB is high.

As a consequence, the signal on the SiPM sensor would be due to both the Cherenkov sources and night sky background emissions. To attenuate, as much as possible, the contribution coming from the latter, it has been proposed to introduce an infrared (IR) filtering window into the optical path. The IR filter has been designed to cut the signal at wave-length above 550 nm (at which the background begins to increase) and would be placed in front of the detectors, working both as filter and as protective window for the SiPM sensors.

The IR filter was implemented as a series of three double coated substrates. This solution allows us to reach high contrast, limiting the number of deposited layers. The substrates were realized as circular Spectrosil foils with a diameter of 400-mm and thickness of 1.5 mm. Each coating is realized as $ZrO_2$-$MgF_2$ multilayer.

To test the filtering efficiency of the window a demonstrator was realized. Three small Spectrosil samples were double coated and assembled as in the ASTRI filtering window configuration. The demonstrator filter transmittance was measured in the 330 – 930 nm spectral range with steps of 50 nm. Moreover, since in the ASTRI telescope configuration the detector is hit by photons coming from a wide range of incidence angles, the measurement was repeated for incidence angles from 10 degrees to 50 degrees. The transmittance curves measured for the different incidence angles and the one corresponding to the calculated mean incidence angle on the filter (44 degrees) are shown in Figure 15. A drop off of about 70% is achieved at wave-lengths greater than 700 nm.

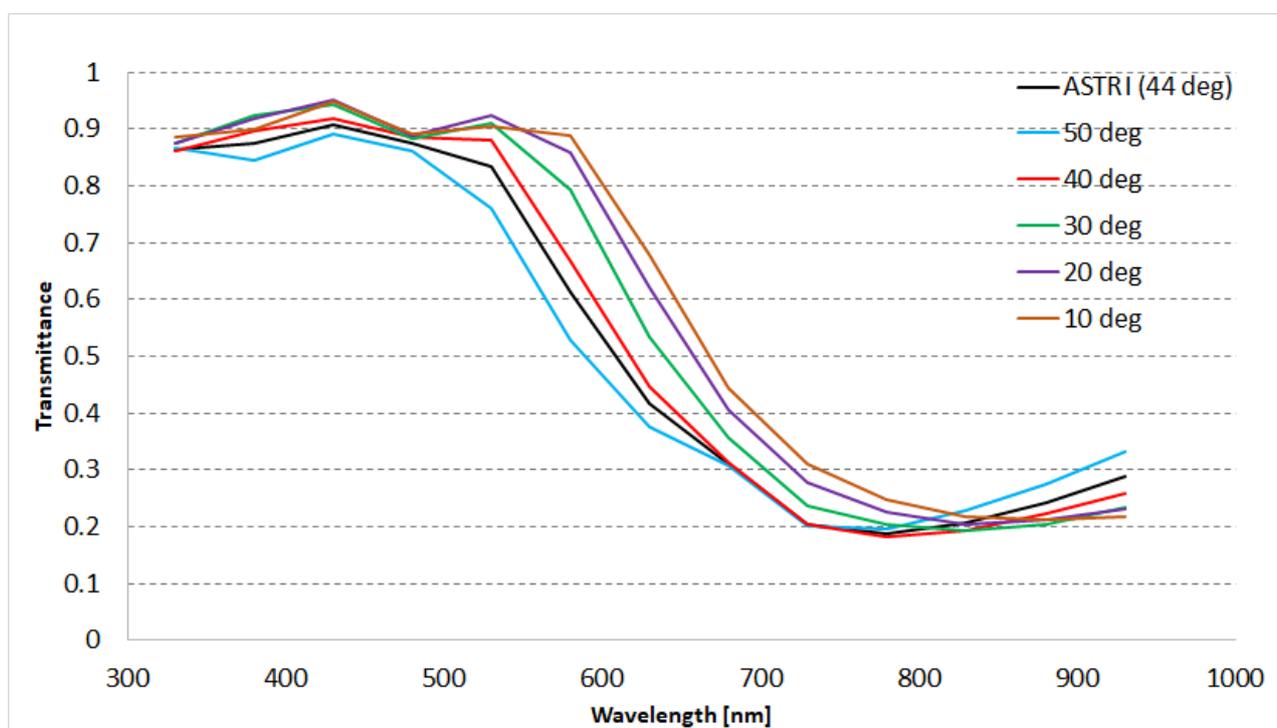

Figure 15. Transmittance in the 300 – 900 nm spectral range at various incident angles of the ASTRI IR filter.

As discussed in section 5, the OCT not only depends on the internal detector features (i.e trenches) and operating conditions (over-voltage) but also on the coating on top of the sensitive area that is responsible for the photons reflection. By extrapolating this behavior to an optics placed close to the detector, the OCT can be expected to increase. Furthermore, a dependence of the distance from the SiPM has to be taken into account, because, at certain distances, reflection could be negligible: the smaller the distance, the greater the OCT. Figure 16 depicts the tiles placement inside the ASTRI camera.

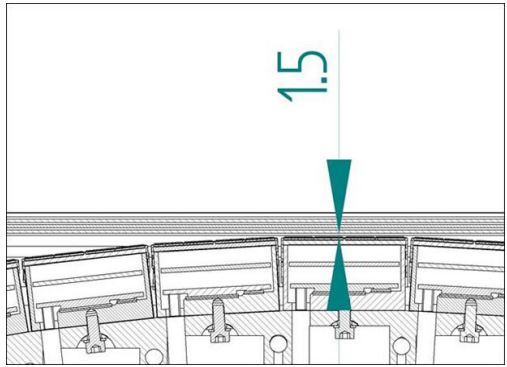

Figure 16. Minimum distance between extrados and intrados of the first sheet of Spectrosil.

As it can be noted the SiPM distance from the filter reaches its minimum of 1.5 mm in the central part of the focal plane. Moreover, the curved focal plane implies a non-constant distance of the pixels from the filter and a rising OCT from the peripheral to the central pixels has to be taken into account.

We used a small reproduction of the filter and by means of appropriate mechanical supports and spacers specially designed for the measurement, we were able to place the sample at 1 mm, 4 mm and 7 mm away from the detector.

Figure 17 shows the measured OCT versus the over-voltage of the LVR2 7075 CN at the mentioned distances from the detector. For comparison, the OCT obtained without the filter is also reported.

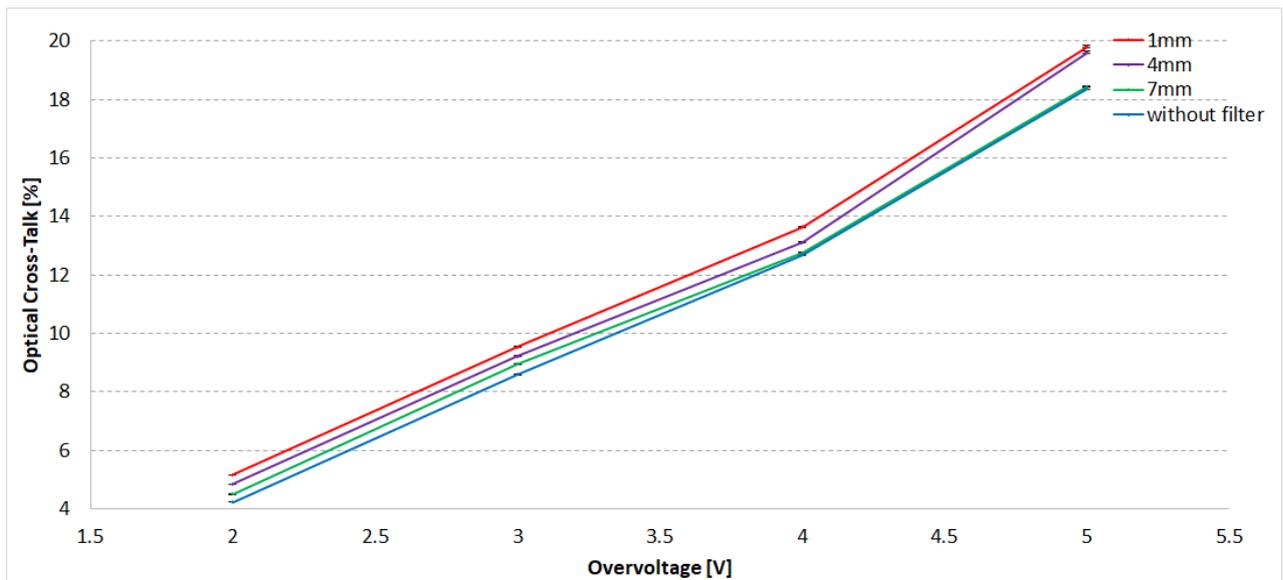

Figure 17. Measured OCT versus over-voltage of the LVR2 7075 CN at distances of 1 mm, 4 mm and 7 mm from the detector. For comparison, we report also the OCT measured without the filter.

It is evident that the OCT decreases with increase in distance between the filter and the SiPM. At distances greater or equal to 7 mm the filter effect is negligible on the OCT increase. In particular, the measurement carried out at an over-voltage of 3.0V shows that, at a distance of 1 mm and of 4 mm, the OCT increases by about 1.0% and 0.6%, respectively. The OCT doesn't change when the filter is 7 mm away from the sensitive surface. The LVR2 7075 CN behaves the same way. Thus we can conclude that to avoid the influence of the filter on the OCT, the filter has to be placed at a distance greater than 6 mm.

## 7. Conclusions and Outlook

In this paper, measurement results of newly available large-area SiPM detectors are reported and discussed. The new S14520 LVR2 series achieve a significant optical cross-talk reduction compared to the LCT5 series due to optical trench optimization and higher photon detection efficiency due to geometrical fill-factor enhancement. In addition, we demonstrated that the S14520 LVR3 series offer improved photon detection capabilities in the near ultraviolet (NUV) spectral region, thus contributing to making the 7 × 7 mm$^2$ with 75 µm micro-cell SiPM detector particularly suitable for Dual-Mirror Small-Sized Telescopes of the CTA project. Measurement results show that, even for large area devices, promising performance in terms of OCT and PDE is achieved. Issues of the OCT measurements have been discussed in great detail. The measurements have shown that the dark pile-up as well as the OCT are affected by the readout electronics if not properly adapted. Finally, the investigation of the use of an optics (in the specific case an IR filter) in front of the device has been also reported. The measurements results demonstrated an OCT increase when the SiPM detector is close to the optics.

**Acknowledgements**

This work is supported by the Italian Ministry of Education, University, and Research (MIUR) with funds specifically assigned to the Italian National Institute of Astrophysics (INAF) for the Cherenkov Telescope Array (CTA), and by the Italian Ministry of Economic Development (MISE) within the "Astronomia Industriale" program. We acknowledge support from the Brazilian Funding Agency FAPESP (Grant 2013/10559-5) and from the South African Department of Science and Technology through Funding Agreement 0227/2014 for the South African Gamma-Ray Astronomy Programme. We gratefully acknowledge financial support from the agencies and organizations listed here: http://www.cta-observatory.org/consortium_acknowledgments.
This paper has gone through internal review by the CTA Consortium.